
\input phyzzx
\def\cL{{\cal L}}
\def\dplus{=\hskip-5pt \raise 0.7pt\hbox{${}_\vert$} ^{\phantom 7}}
\def\dplusup{=\hskip-5.1pt \raise 5.4pt\hbox{${}_\vert$} ^{\phantom 7}}

\def\dplus{=\hskip-4.8pt \raise 0.7pt\hbox{${}_\vert$} ^{\phantom 7}}

\def\pmb#1{\setbox0=\hbox{#1} \kern-.025em\copy0\kern-\wd0
\kern0.05em\copy0\kern-\wd0 \kern-.025em\raise.0433em\box0}

\def\cE{{\cal E}}

\def\cM{{\cal M}}
\def\cL{{\cal L}}

\REF\hls{R. Haag, J. Lopuszanski, and M. Sohnius, Nucl. Phys. {\bf B88} (1975)
257.}
\REF\agf{ L. Alvarez-Gaum{\' e} and D.Z. Freedman, Commun. Math. Phys.
{\bf 80} (1981), 443.}
\REF\ghr{S.J. Gates, C.M. Hull and M. Ro{\v c}ek, Nucl. Phys. {\bf B248}
(1984) 157; C.M. Hull, {\it Super Field Theories} ed H. Lee and G. Kunstatter
(New York: Plenum) (1986).}
\REF\hpa {P.S. Howe and G. Papadopoulos, Nucl .Phys. {\bf B289} (1987) 264;
Class. Quantum Grav. {\bf 5} (1988) 1647.}
\REF\mr{M. Ro{\v c}ek, K. Shoutens and A. Sevrin, Phys. Lett. {\bf 265B} (1991)
303.
\break
B. Kim and M. Ro{\v c}ek, ``Complex structures, duality and WZW models in
extended superspace",
hep-th 9406063. }
\REF\sev {Ph. Spindel, A. Sevrin, W. Troost and A. Van Proeyen, Nucl. Phys.
{\bf B308} (1988) 662;
{\bf B311} (1988) 465.}
\REF\hw{C.M. Hull and E. Witten, Phys. Lett. {\bf 160B} (1985) 398. }
\REF\nr{B. deWit and P. van Nieuwenhuizen, Nucl. Phys. {\bf B312} (1989) 58.}
\REF\nra{G.W. Delius, M.Ro{\v c}ek, A. Sevrin and P. van Nieuwenhuizen, Nucl.
Phys. {\bf
B324} (1989) 523.}
\REF\hpb{P.S. Howe and G. Papadopoulos, Commun. Math. Phys. {\bf 151} (1993)
467.}
\REF\gp{G. Papadopoulos, Phys. Lett. {\bf 238B} (1990) 75; Commun. Math. Phys.
{\bf 144} (1992)
491.}
 \REF\sen {A. Sen, Nucl. Phys. {\bf B278} (1986) 289.}
\REF\hpc{P.S. Howe and G. Papadopoulos, Class. Quantum. Grav. {\bf 4} (1987)
1749.}
\REF\hps{P.S. Howe, G. Papadopoulos and K.S. Stelle, Nucl. Phys. {\bf B296}
(1988) 26.}
\REF\bz {B. Zumino, {\sl Relativity Groups and Topology II}, ed. B.S. DeWitt
and R. Stora
(Amsterdam: North Holland).}
\REF\ht{C.M. Hull and P. Townsend, Phys. Lett. {\bf 178B} (1986) 187. }
\REF\nel {G. Moore and P. Nelson, Commun. Math. Phys. {\bf 100} (1985) 83.}
\REF\hpt{C.M. Hull, G. Papadopoulos and P.K. Townsend, Phys. Lett.
{\bf 316B} (1993) 291.}
\REF\pt{G. Papadopoulos and P.K. Townsend, Class. Quantum Grav. {\bf 11} (1994)
515; {\bf 11}
(1994) 2163.}
\REF\witten {E. Witten, Commun. Math. Phys. {\bf 118} (1988) 411.}
\REF\pkt {J. A. de Azc{\'a}rraga, J. P. Gauntlett, J. M. Izquierdo, and P.K.
Townsend, Phys. Rev.
Lett. {\bf 63} 22 (1989) 2443.}


\Pubnum{ \vbox{ \hbox{R/95/3} \hbox{}} }
\pubtype{}
\date{February 15, 1995}

\titlepage

\title{(2,0)-supersymmetric sigma models and almost complex structures}

\author{G. Papadopoulos}
\address{D.A.M.T.P
 \break University of Cambridge\break
         Silver Street \break Cambridge CB3 9EW}

\abstract {We find a new class of (2,0)-supersymmetric two-dimensional sigma
models with
torsion and target spaces almost complex manifolds extending similar results
for models with
(2,2) supersymmetry.  These models are invariant under a new symmetry which is
generated by a Noether charge of Lorentz weight one and it is associated to the
Nijenhuis tensor of the almost complex structure of the sigma model target
manifold.  We compute
the Poisson bracket algebra of charges of the above  (2,0)-and
(2,2)-supersymmetric sigma models
and show that it closes  but it is not isomorphic to the
standard (2,0) and (2,2) supersymmetry algebra, respectively.  Examples of such
(2,0)- and
(2,2)-supersymmetric sigma models with target spaces group manifolds are also
given.  In
addition, we study the quantisation of the (2,0)-supersymmetric sigma models,
compute the
anomalies of their classical symmetries and examine their cancellation.
Furthermore, we examine
the massive extension of (2,0)-supersymmetric sigma models with target spaces
almost complex
manifolds, and study the topological twist of the new  supersymmetry algebras.}
\vskip 0.5cm

\endpage

\pagenumber=2



\def\RN{{\cal R}}
\def\CN{{\cal C}}

\def\dim{{\rm dim}}

\def\cE {{\cal {E}}}

\def\fff{\vrule width0.5pt height5pt depth1pt}
\def\pp{{{ =\hskip-3.75pt{\fff}}\hskip3.75pt }}

\def\cM {{\cal{M}}}



\chapter{Introduction}

The (p,q) supersymmetry algebra in two dimensions is
$$
\{S_+^I,S_+^J\}=2\delta^{IJ} T_\pp, \qquad \{S_-^{I'},S_-^{J'}\}=2\delta^{I'J'}
T_=\ ,
\qquad
\{S_+^I,S_-^{I'}\}=Z^{II'}\ ,
\eqn\inone
$$
where $\{S_+^I; I=0,\dots, p-1\}$ are the `left' supersymmetry charges,
$\{S_-^{I'}; I'=0,\dots,
q-1\}$ are the `right' supersymmetry charges, $T_\pp=E+P$, $T_==E-P$, $E$ is
the energy,  $P$
is the momentum and $Z^{II'}$ are central charges of the algebra\foot{ This
algebra can be
enhanced by adding additional generators that rotate the supersymmetry charges.
This
possibility will not be considered here.}. The subscripts in the above  charges
denote
the Lorentz weight of the charge; for example $S_+^I$ has Lorentz weight
${1\over2}$,
$S_-^{I'}$ has Lorentz weight $-{1\over2}$ and $Z^{II'}$ has Lorentz weight
zero.
The  central charges are zero for massless theories, {\sl i.e} theories without
a parameter with
dimension that of a mass. In the following, we will discuss only {\sl massless}
models unless it is otherwise stated. The supersymmetry algebra
\inone, up to an isomorphism, is the one expected for a supersymmetric theory
from the
Haag-Lopuszanski-Sohnius theorem [\hls].

Realisations of the supersymmetry algebra \inone\ in terms of Poisson bracket
algebras of  charges
of supersymmetric sigma models in two dimensions  have been extensively studied
in the
literature [\agf, \ghr,\hpa].  The main observation regarding these
realisations is that (p,q)
supersymmetry imposes restrictions on the geometry of the sigma model
manifolds.  To illustrate
this,  we will  summarise the geometry of the target spaces of massless (2,2)-
and
(2,0)-supersymmetric sigma models.  The geometry of the target space
$\cM$ of two-dimensional supersymmetric sigma models with (2,2) supersymmetry
depends upon the
properties of two (1,1) tensors $I$ and $J$ on $\cM$ which appear naturally in
the (2,0) and (0,2)
supersymmetry transformations of the fields.   For (2,2)-supersymmetric sigma
models  without
Wess-Zumino term, or `torsion',
$I=J$, $I$ is a complex structure which is covariantly constant with respect to
the
Levi-Civita connection $\Gamma(g)$ of the sigma model metric $g$ and $g$ is
hermitian with
respect to $I$, {\sl i.e.} the sigma model manifold $\cM$ is K\"ahler [\agf].
These models admit a
conventional superfield formulation in terms of chiral superfields. For
(2,2)-supersymmetric sigma models with Wess-Zumino term $b$ and on-shell
closure of the algebra
of (2,2) supersymmetry transformations, $I$ and
$J$ are again complex structures, but $I\not=J$, the sigma model metric $g$ is
hermitian
with respect to both complex structures, and $I$ and $J$ are covariantly
constant with
respect to the connections $\Gamma^{(+)}\equiv \Gamma(g)+H$ and
$\Gamma^{(-)}\equiv
\Gamma(g)-H$ with torsion, respectively; the torsion $H={3\over2} db$.   A
superfield
formulation for some of these models has been proposed in ref. [\mr].  Examples
of sigma models
with on-shell (2,2) supersymmetry are the supersymmetric extensions of the WZW
models [\sev].  The
algebra of supersymmetry transformations of (2,2) sigma models with torsion
closes off-shell
provided that, in addition to the conditions for on-shell closure mentioned
above, the complex
structures $I$ and $J$ commute, $IJ=JI$.  In this case there is a conventional
(2,2) superfield formulation of the theory [\ghr].

Sigma models with (2,0) supersymmetry are naturally associated with a pair of a
Riemannian manifold $\cM$, the sigma model manifold, and a vector bundle $\cE$
over $\cM$.
The geometry of $\cM$ and $\cE$ of (2,0)-supersymmetric sigma models with
on-shell
supersymmetry also depends upon the properties of a (1,1) tensor $I$ on $\cM$.
On-shell
closure of the algebra of (2,0) supersymmetry transformations  requires that
the tensor $I$
is a complex structure which is covariantly constant with respect to
$\Gamma^{(+)}$
connection, the sigma model metric is hermitian with respect to $I$ and the
curvature of
a connection of the $\cE$ bundle is an (1,1) tensor with respect to $I$ [\hw],
{\sl i.e.} $\cM$
is a hermitian manifold and the complexified bundle of $\cE$ is holomorphic.
For
off-shell closure of the algebra of (2,0) supersymmetry transformations, in
addition to
$I$ and the conditions mentioned above for on-shell closure of the (2,0)
supersymmetry transformations, a new (1,1) tensor
$\hat{I}$ is required  on the fibre of the vector bundle
$\cE$ such that $\hat{I}$ is  a complex structure, the bundle space of $\cE$ is
a complex
manifold with respect to the pair of complex structures $(I, \hat{I})$ and a
fibre metric of
$\cE$ is hermitian with respect to $\hat{I}$ [\hpa].  The (2,0)-supersymmetric
sigma models
with off-shell supersymmetry admit a conventional superfield formulation in
terms
of constrained superfields [\hpa].

More recently, new (2,2)-supersymmetric sigma models with torsion have been
considered in
refs.[\nr,\nra,\hpb].  The tensors $I$ and $J$ of these models are {\it almost}
complex
structures that are covariantly constant with respect to the connections
$\Gamma^{(+)}$ and
$\Gamma^{(-)}$, respectively, and the sigma model metric is hermitian with
respect to both
almost complex structures $I$ and $J$.  These new models differ from the
standard (2,2)
supersymmetric sigma models mentioned above because their action is invariant
under
transformations generated by the Nijenhuis tensors of the almost complex
structures $I$ and
$J$; we will call these symmetries Nijenhuis symmetries.  The algebra of
supersymmetry and
Nijenhuis transformations of the above models closes [\hpb] provided that their
parameters are
constant. Note that if the torsion
$H$ is zero the Nijenhuis tensors of $I$ and $J$ are zero as well and the sigma
model manifold is
K{\" a}hler.

Our main interest in this paper is to find the conditions for the existence of
sigma
models with (2,0) supersymmetry and target spaces which are {\sl not}  complex
manifolds.   We
will show that such models have target spaces almost complex manifolds and new
symmetries
associated with the Nijenhuis tensor of the almost complex structures.  We will
compute the Poisson
bracket algebra of the charges of the (2,0)-supersymmetric sigma models with
target space almost
complex manifolds and we will show that the non-vanishing Poisson brackets are
the following:
$$
\{S^{{}^0}_+, S^{{}^0}_+\}=2T_\pp, \qquad \{S^{1}_+, S^{1}_+\}=2(T_\pp+N_\pp) \
,
\eqn\intwo
$$
where the charges $S^{{}^0}_+$, $S^1_+$, $T_\pp$ are as in \inone\ and  $N_\pp$
is the charge that
generates the Nijenhuis symmetry of the model.  We will also compute the
Poisson bracket
algebra of charges of (2,2)-supersymmetric sigma models with target spaces
almost complex
manifolds and show that it is isomorphic to two commuting copies of \intwo. The
supersymmetry
algebra \intwo\ of the above (2,0)-supersymmetric sigma models is {\sl not}
isomorphic to the
corresponding standard supersymmetry algebra \inone.  We will give examples of
(2,2)- and
(2,0)-supersymmetric sigma models with Nijenhuis symmetries and  target spaces
group manifolds.
Furthermore, we will also study the quantisation of (2,0)-supersymmetric sigma
models and compute
the anomalies associated with the (2,0) supersymmetry and Nijenhuis
transformations
using the descent equations. We will also examine the cancellation of these
anomalies and
we will show that  some of (2,0))-supersymmetric sigma models with classical
Nijenhuis symmetries
and target spaces group manifolds are anomaly free.  In addition, we will
extend the above
results to massive (2,0)-supersymmetric sigma models and we will examine a
topological twist
of the supersymmetry algebra \intwo.

This paper has been organised as follows: In section two, the
(2,2)-supersymmetric sigma models
with target spaces almost complex manifolds will be reviewed.  In section
three, the new
(2,0)-supersymmetric sigma  models with Nijenhuis symmetries, will be
presented. In section
four, examples of (2,2)- and (2,0)-supersymmetric sigma models with Nijenhuis
symmetries and
target spaces group manifolds will be given.  In section five, the Poisson
bracket algebra
of the charges of (2,2)- and (2,0)-supersymmetric sigma models with Nijenhuis
symmetries will be
presented.  In section six, the anomalies in the classical symmetries of the
(2,0)-supersymmetric sigma model will be examined.  In
section seven, the  massive extension of (2,0)- and (2,2)-supersymmetric sigma
models with
Nijenhuis symmetries and a topological twisting of the supersymmetry algebra
with Nijenhuis
charges  will briefly be examined.  A summary will be given in section eight.


\chapter{The (2,2) model}

Let $\cM$ be a Riemannian manifold with metric $g$ and a locally defined
two-form $b$.  The
patching condition for $b$ is $b'=b+dm$ where $m$ is an one-form defined on the
intersection of any two open sets of $\cM$.    The action of the
(1,1)-supersymmetric model is
$$
I=\int d^2x d\theta^+ d\theta^- (g+b)_{ij} D_-\phi^iD_+\phi^j\ ,
\eqn\aone
$$
where $(x^\pp,x^=,\theta^+,\theta^-)$ are the co-ordinates of (1,1) superspace,
$\Xi^{(1,1)}$, $(x^\pp,x^=)=(x+t,t-x)$) are light-cone co-ordinates, the
indices $i,j=1,\dots,
\rm{dim}\ \cM$, $\phi$ is a (1,1)-superfield that takes values in $\cM$, and
$D_-, D_+$ are the
supersymmetry derivatives of (1,1) superspace, {\sl i.e.}
$$
D_+^2=i\partial_\pp\ , \qquad  D_-^2=i\partial_= \ .
\eqn\atwo
$$

The (2,0) and (0,2) supersymmetry transformations can be written in terms of
(1,1)
superfields as follows:
$$\eqalign{
\delta_I\phi^i&= a_- I^i_jD_+\phi^j\ ,
\cr
\delta_J\phi^i&= a_+ J^i_jD_-\phi^j\ ,}
\eqn\athree
$$
where $I$ and $J$ are (1,1) tensors on the sigma model manifold $\cM$ and
$a_+$, $a_-$ are
the constant anti-commuting parameters of the transformations.  These
transformations leave
the action \aone\ invariant provided that
$$\eqalign{
\nabla^{(+)}_i I^j{}_k&=0\ , \qquad \nabla^{(-)}_i J^j{}_k=0\ ,
\cr
g_{k(i} I^k{}_{j)}&=0\ , \qquad g_{k(i} J^k{}_{j)}=0\ ,}
\eqn\afour
$$
where
$$
\Gamma^{(\pm)}{}^i_{jk}=\{^i_{jk}\}\pm H^i{}_{jk}\ ,
\eqn\afive
$$
and
$$
H_{ijk}={3\over 2} \partial_{[i} b_{jk]}\ .
\eqn\asix
$$

The algebra of transformations of eqn.\athree\ closes on-shell as follows:
$$\eqalign{
[\delta_I,\delta'_I]\phi^i&= \delta_{N(I)}\phi^i+2 i a'_- a_-
\partial_\pp\phi^i\ ,
\cr
[\delta_J,\delta'_J]\phi^i&= \delta_{N(J)}\phi^i+2 i a'_+ a_+
\partial_=\phi^i\ ,
\cr
[\delta_I,\delta_J]\phi^i&=0\ ,}
\eqn\aseven
$$
provided that
$$
I^2=-1, \qquad J^2=-1\ ,
\eqn\aeight
$$
where
$$\eqalign{
\delta_{N(I)}\phi^i&\equiv a_=  N(I)^i{}_{jk} D_+\phi^j  D_+\phi^k\ ,
\cr
\delta_{N(J)}\phi^i&\equiv a_\pp N(J)^i{}_{jk} D_-\phi^j D_-\phi^k \ , }
\eqn\anine
$$
$a_=$,  $a_\pp$ are the parameters of the transformations ($a_==a'_- a_-$,
$a_\pp=a'_+ a_+$ in the commutator \aseven), and $N(I)$ and $N(J)$ are the
Nijenhuis tensors of $I$
and
$J$, respectively.  The Nijenhuis tensor of a (1,1) tensor $I$ on $\cM$ is
$$
N(I)^i{}_{jk}= 2\big(I^m{}_{[j}\partial_{|m|} I^i{}_{k]}-I^i{}_m\partial_{[j}
I^m{}_{k]}\big)\ .
\eqn\aten
$$

The transformations \anine\ are symmetries of the action \aone\ because they
appear in the
commutator \aseven\ of two symmetries of the theory together with the
translations. Note that the
translations are by themselves symmetries of the action \aone.   We can also
verify this by a
straightforward calculation using the following properties of the Nijenhuis
tensor of the almost
complex structures $I$ and $J$:
$$
\nabla^{(+)}_iN(I)_{jkl}=0\ , \qquad \nabla^{(-)}_iN(J)_{jkl}=0\ ,
\eqn\aeleven
$$
and
$$
N(I)_{ijk}=N(I)_{[ijk]}\ , \qquad  N(J)_{ijk}=N(J)_{[ijk]}\ .
\eqn\atwelve
$$
The commutator of Nijenhuis transformations \anine\ with themselves, and with
the (2,0) and (0,2) supersymmetry transformations given in \athree\ vanishes
[\hpb].  Note that
$$
N(I)_{kij} I^k{}_m+(m,i)=0, \qquad N(J)_{kij} J^k{}_m+(m,i)=0\ .
\eqn\athirteen
$$
Finally, it is worth pointing out though that, in
contradistinction to the case of (2,2)-supersymmetric sigma models with target
spaces complex
manifolds, the parameters of the
transformations \aseven\ cannot be promoted to semi-local ones because the
algebra of supersymmetry
transformations \athree\ and Nijenhuis symmetries \anine\ does not close.

\chapter{The (2,0) models}

Let $\cM$ be a Riemannian manifold with metric $g$ and a locally defined 2-form
$b$ as in the
previous section, and
$\cE$ be a vector bundle over $\cM$ with connection $A$ ($\rm{rank}\,\cE=k$)
and fibre
metric $h$.  We choose the connection $A$ such that $\nabla_ih=0$ \foot{Note
that given a
connection $A$ of $\cE$ with fibre metric $h$ there always exist another
connection of $\cE$ with
respect to which $h$ is covariantly constant.}. The action of
(1,0)-supersymmetric sigma model is
$$
I=-i\int\! d^2 x d \theta ^+ \big\{ (g_{ij} +b_{ij})  D_+\phi^i\partial_=\phi^j
+ih_{ab}\; \psi_-^a\nabla_+\psi_-^b  \big\} \ ,
\eqn\bone
$$
where  $(x^\pp,x^=, \theta^+)$ are the co-ordinates of (1,0) superspace,
$\Xi^{(1,0)}$,
$(x^\pp,x^=)=(x+t, t-x)$ are light-cone co-ordinates, the indices $a,b=1,
\dots, k$,  $D_+$ is the
supersymmetry derivative
($D_+^2=i \partial_\pp$) and
$$
\nabla_+\psi_-^b\equiv (D_+\psi_-^b + D_+\phi^i A_{i}{}^b{}_c\psi_-^c)\ .
\eqn\btwo
$$
The fields of (1,0) supersymmetric sigma model are the following: the scalar
superfield
$\phi(x,\theta^+)$ which is a map from the (1,0) superspace, $\Xi^{(1,0)}$,
into the
target manifold ${\cal M}$, and the spinor superfield $\psi_-^a(x,\theta ^+)$
which is a
section  of the vector bundle $\xi_-\otimes \phi^*\cE$ where $\xi_-$ is the
spin bundle over
$\Xi^{(1,0)}$.  The part of \bone\ that contains the $\psi$ field is called
either the fermionic or
Yang-Mills sector of the sigma model action.

To find sigma models with (2,0) supersymmetry, we introduce the transformations
$$\eqalign{
\delta_I\phi^i&=a_- I^i{}_j D_+\phi^j
\cr
\delta_I\psi_-^a&=-A_i{}^a{}_b \delta_I\phi^i \psi_-^b+ a_- \hat{I}^a{}_b
\nabla_+\psi^b_-\ .}
\eqn\bthree
$$
written in terms of (1,0) superfields, where $I$ is a (1,1) tensor on $\cM$,
$\hat{I}$
is a (1,1) tensor on the fibre of $\cE$ and $a_-$ is the constant
anti-commuting parameter of
the transformations.  The commutator of these transformations on the field
$\phi$ is the same as the one given in the previous section for the $\delta_I$
transformations on
$\phi$ for the case (2,2)-supersymmetric sigma models, {\sl i.e.} the
commutator on
$\phi$ of the transformations \bthree\ closes to translations and Nijenhuis
transformations
provided that $I$ is and almost complex structure ($I^2=-1$).  The commutator
on the field
$\psi$ is
$$\eqalign{
[\delta_I,\delta'_I]\psi_-^a=-&A_i{}^a{}_b [\delta_I,\delta'_I]\phi^i \psi_-^b
\cr
-&a'_- a_- (F_{kl}{}^a{}_b I^k{}_i I^l{}_j- F_{ij}{}^a{}_b) D_+\phi^i D_+\phi^j
\psi_-^b
\cr
+&2a'_- a_- (\nabla_j\hat{I}^a{}_b I^j{}_i- \hat{I}^a{}_c\nabla_i\hat{I}^c{}_b)
D_+\phi^i \nabla_+\psi_-^b
\cr
+& 2 i a'_- a_- \nabla_\pp\psi^a\ ,}
 \eqn\bfour
$$
where
$$
F_{ij}{}^a{}_b= \partial_i A_j^a{}_b-\partial_j
A_i^a{}_b+A_i^a{}_cA_j^c{}_b-A_j^a{}_cA_i^c{}_b
\eqn\bfoura
$$
is the curvature of the connection $A$.
There are two cases to consider the following: Case (i), the commutator on
$\psi$ closes {\sl
on-shell} to translations and to the Nijenhuis transformations
$$\eqalign{
\delta_{N}\phi^i&=a_= N^i_{jk} D_+\phi^j D_+\phi^k\ ,
\cr
\delta_{N}\psi_-^a&=- A_i{}^a{}_b \delta_{N}\phi^i \psi_-^b\ ,}
\eqn\bfive
$$
where $a_=$ is the parameter of the  Nijenhuis transformations ($a_==a'_- a_-$
in the commutator
\bfour), provided that
$$
F_{kl}{}^a{}_b I^k{}_i I^l{}_j- F_{ij}{}^a{}_b=0\ .
\eqn\bsix
$$
Case (ii), the commutator \bfour\ closes {\sl off-shell} to translations and
the Nijenhuis
symmetry \bfive, provided that in addition to \bsix, the conditions
$$
\hat{I}^2=-1, \qquad \nabla_j\hat{I}^a{}_b I^j{}_i-
\hat{I}^a{}_c\nabla_i\hat{I}^c{}_b=0
\eqn\bseven
$$
are satisfied. Therefore ${\hat I}$ is an almost complex structure on the fibre
of $\cE$. It is
worth pointing out that given a connection $A$ of $\cE$,  we can always choose
another
connection ${\tilde A}=A -{1\over2} {\hat I}\nabla {\hat I}$ on $\cE$ such that
${\tilde\nabla}_i \hat {I}=0$ and ${\tilde\nabla}h=0$, if $\nabla h=0$, where
$h$ is the fibre
metric of $\cE$.

The commutator of the Nijenhuis \bfive\ with the (2,0) supersymmetry
transformations on both fields $\phi$ and $\psi$ vanishes, {\sl i.e.}
$$
[\delta_I, \delta_{N}]\phi^i=0, \qquad [\delta_I, \delta_{N}]\psi^a_-=0\ .
\eqn\beleven
$$
It is straightforward to verify this on the
field $\phi$ because this commutator is the same as the one of the Nijenhuis
with (2,0)
supersymmetry transformations in the case of (2,2) models reviewed in section
two.   To examine the commutator of the Nijenhuis with the (2,0) supersymmetry
transformations
on the field $\psi$, there are two cases to consider the following:  Case (i),
the commutator on
$\psi$ vanishes {\sl on-shell} provided that
$$
 F_{mn}{}^a{}_b I^m{}_{[i} N^n{}_{jk]}=0\ .
\eqn\beight
$$
We can show that \beight\ is not an independent condition and it can be derived
from the conditions
that $I$ is an almost complex structure and $F$ satisfies \bsix\ using the
Bianchi identities. We
can also show that
$$
 N^m{}_{[ij} F_{k]m}{}^a{}_b =0\ .
\eqn\bnine
$$
Case (ii), the commutator
on $\psi$ vanishes {\sl off-shell} provided that, in addition to \beight,
$$
N(I)^i{}_{jk} {}\nabla_i \hat{I}=0\ .
\eqn\bten
$$
The commutator of two Nijenhuis symmetries vanishes, as well, without further
conditions.

The action is invariant under both (2,0) supersymmetry and Nijenhuis
transformations provided that, in addition to the conditions obtained above for
the closure
of the algebra of these transformations, the following conditions are
satisfied:
$$
\nabla^{(+)}_i I^j{}_k=0, \qquad g_{k(i} I^k{}_{j)}=0\, \qquad h_{c(a} {\hat
I}^c{}_{b)}=0\  .
\eqn\btwelve
$$

To summarise, the independent conditions for the invariance of the action and
the {\sl on-shell}
closure of the algebra of supersymmetry and Nijenhuis transformations of
(2,0)-supersymmetric
sigma models are the following: the first equation in \aeight, \bsix\ and
\btwelve. Note that in this case we can set ${\hat I}=0$. The independent
conditions for the
invariance of the action and the {\sl off-shell} closure of the algebra of
supersymmetry and
Nijenhuis transformations are the
following: the first equation in
\aeight,
\bsix, \bseven, \bten\  and \btwelve.
Therefore, the target manifolds of sigma models with {\sl on-shell} (2,0)
supersymmetry and Nijenhuis symmetry are almost complex manifolds, the almost
complex
structure $I$ is covariantly constant with respect to the $\Gamma^{(+)}$
connection, i.e. the
holonomy of
$\Gamma^{(+)}$ is a subgroup of $U(m)$ ($\rm {dim} \cM=2m$), the metric $g$ is
hermitian with respect to $I$ and the curvature of the connection $A$ of the
bundle $\cE$ is an (1,1) form with respect to $I$.  In addition to
the above restrictions on the geometry of the target manifold $\cM$ and vector
bundle $\cE$
required by on-shell (2,0) supersymmetry, the bundle
$\cE$ of sigma models with  {\sl off-shell} (2,0) supersymmetry and Nijenhuis
symmetry must
admit a fibre almost complex structure
${\hat I}$, i.e. the structure group of $\cE$ is a subgroup of $U({k\over2})$,
and the fibre
metric $h$ is hermitian with respect to ${\hat I}$.


\chapter{Models on group manifolds}

Some explicit examples of models with (2,2) or (2,0) supersymmetry and
Nijenhuis symmetry
are found by consideration of sigma models with a group manifold as their
target space.  Let
$K$ be a group manifold with Lie algebra $\cL(K)$. The left $L^A$ and right
$R^A$ invariant
frames on $K$ are defined as follows:
$$
k^{-1}dk=L^A t_A\qquad dk\, k^{-1}=R^A t_A, \qquad k\in K\ ,
\eqn\seone
$$
where $\{t_A\}$ is a basis in $\cL(K)$,  $[t_A,t_B]=f_{AB}{}^Ct_C$, the indices
$A,B,C=1,
\dots, \rm{dim} \cL(K)$, and
$f_{AB}{}^C$ are the structure constants of $\cL(K)$.  The Maurer-Cartan
equations are
$$
dL^A=-{1\over2} f_{BC}{}^AL^BL^C\qquad  dR^A={1\over2} f_{BC}{}^AR^BR^C\ .
\eqn\setwo
$$
The sigma model metric $g$ and torsion $H$ are chosen to be the bi-invariant
tensors
$$
\eqalign{
g_{ij}&=\kappa_{AB} L^A_i L^B_j=\kappa_{AB} R^A_i R^B_j\ ,
\cr
H_{ijk}&=-{\lambda\over 2} f_{ABC}L^A_i L^B_j L^C_k=-{\lambda\over 2}
f_{ABC}R^A_i R^B_j
R^C_k\ ,}
\eqn\sethree
$$
where $\kappa_{AB}$ is an invariant non-degenerate quadratic form on $\cL(K)$,
$f_{ABC}=f_{AB}{}^D\kappa_{DC}$ and $\lambda$ is a real number.  The
Wess-Zumino-Witten models
that we will consider here are those for which $|\lambda|=1$.  For these values
of
$\lambda$ both $\Gamma^{(+)}$ and
$\Gamma^{(-)}$ connections are flat and the group manifold is parallelizable
with respect
to both connections.  In the following we will choose $\lambda=1$.

We will first consider the sigma models with (2,2) supersymmetry (see also ref.
[\nra]).  The task
is to find solutions to the conditions required by (2,2)-supersymmetry on the
almost complex
structures $I$ and $J$, {\sl i.e.} to find solutions to the conditions of eqns.
\afour. We can solve the first two equations in \afour\ by setting
$$
I^i{}_j= L^i_A I^A{}_B L^B{}_j\ , \qquad J^i{}_j= R^i_A J^A{}_B R^B{}_j\ ,
\eqn\sefour
$$
where the matrices $\{I^A{}_B\}$ and $\{J^A{}_B\}$ are constant
\foot{We use the same symbols $I$ and $J$ to denote the almost complex
structure on $K$ and
their associated constant tensors on $\cL(K)$.  To avoid confusion when we
refer to the latter, we
will use Lie algebra indices.}.
The condition that both
$I^i{}_j$ and
$J^i{}_j$ are almost complex structures implies that
$$
I^A{}_C I^C{}_B=-\delta^A{}_B \qquad  J^A{}_C J^C{}_B=-\delta^A{}_B
\eqn\sefive
$$
The last two equations of \afour\ then become
$$
\kappa_{CD} I^C{}_A I^D{}_B=\kappa_{AB}\qquad  \kappa_{CD} J^C{}_A
J^D{}_B=\kappa_{AB}\ .
\eqn\sesix
$$
The conditions for existence of (2,2) supersymmetric sigma models reduces in
this case to
the algebraic equations \sefive\ and \sesix.  Since one may set
$I^A{}_B=J^A{}_B$,
a group manifold $K$ is the target manifold of a (2,2)-supersymmetric sigma
model with Nijenhuis
symmetries provided that there is a (constant) complex structure $I^A{}_B$ and
an invariant
quadratic form $\kappa$ on the Lie algebra $\cL(K)$ which is Hermitian with
respect to
$I^A{}_B$.  If $K$ is a simple compact Lie group, then the space of independent
parameters
that parameterise classically the different (2,2)-supersymmetric sigma models
with Nijenhuis
symmetry (the moduli space of the theory) is as follows:  First  the constant
positive conformal
factor that scales  quadratic form $\kappa$. This space is topologically
$\RN^+$ and
parameterises the size of
$K$.  Note that every simple Lie group has a unique invariant non-degenerate
quadratic form up to
scaling with a constant conformal factor.  Second, for each metric there are
${SO(2m)\over U(m)}$ complex structures on $\cL(K)$ that satisfy \sesix ($\dim
K=2m$).  So the
moduli space of a (2,2)-supersymmetric sigma model with Nijenhuis symmetries
and target space an
even-dimensional group manifold can be thought as a bundle with base space
$\RN^+$ and fibre
the space ${SO(2m)\over U(m)}\times {SO(2m)\over U(m)}$.  Quantum mechanically
though the coupling
constant,  {\sl i.e} the conformal factor, of the WZW model is quantised and
therefore the space
$\RN^+$ becomes a lattice.  The above results can be easily generalised for any
semisimple group
$K$.

To give examples of sigma models with on-shell (2,0) supersymmetry and target
space group manifolds, we must find solutions to the equation \bsix\ in
addition to those
satisfied by the almost complex structure $I$.  The conditions on the almost
complex structure $I$
required by (2,0) supersymmetry are the same as those required by (2,2)
supersymmetry and therefore
they reduce to the algebraic conditions \sefive\ and  \sesix\ for
$I^A{}_B$ and $\kappa_{AB}$.  For compact simple Lie groups
the space of parameters is again a bundle space with base space $\RN^+$ and
fibre
${SO(2m)\over U(m)}$.      Now it remains to solve the
equation \bsix.  For this we will assume further that the fermionic sector of
the (2,0) sigma
model is invariant under the left action of $K$\foot{For more details on the
symmetries of the
fermionic sector of the sigma model action see ref.[\gp].} If this is the case
the bundle
$\cE$ is topologically trivial and  we can use the standard trivialisation of
$\cE$ to
identify the connections of $\cE$ with the $\cL(H)$-valued one-forms on $K$
where $H$ is the gauge group of the connections $A$.  The left invariant
connections can now be
written as
$$
A_i=L_i^A \omega_A
\eqn\seseven
$$
where  $\omega$ is constant and can be thought as a linear map from
the Lie algebra $\cL(K)$ of $K$ into $\cL(H)$.  The equation \bsix\ is
then equivalent to
$$
(-f^E_{AB} \omega_E^r+ f^r_{st}\omega_A^s \omega_B^t) I^A{}_C I^B{}_D=
(-f^E_{CD}
\omega_E^r+ f^r_{st}\omega_C^s \omega_D^t)\ ,
\eqn\seeight
$$
where $f^r_{st}$ are the structure constants  $\cL(H)$ and
$r,s,t=1, \dots, {\rm dim}\,\cL(H)$. Using the complex structure $I^A{}_B$ to
decompose
$\cL(K)\otimes \CN$ into holomorphic $\cL(K)^{(1,0)}$ and anti-holomorphic
$\cL(K)^{(0,1)}$
subspaces, we can rewrite the equation
\seeight\ as follows:
$$
-f^E_{\alpha\beta} \omega_E^r+ f^r_{st}\omega_{\alpha}^s \omega_{\beta}^t=0\ ,
\qquad
-f^E_{{\bar \alpha}{\bar \beta}} \omega_E^r+ f^r_{st}\omega_{{\bar \alpha}}^s
\omega_{{\bar
\beta}}^t=0\ ,
\eqn\senine
$$
where the indices $A=(\alpha, \bar {\alpha})$ and $\alpha,
\beta=1,\cdots,m$.  Further simplification of \senine\ does not seem possible
for the case
that $I$ is an almost complex structure.  Observe though that $\omega=0$ is a
solution of
\senine. However if the Nijenhuis tensor of
$I$ is zero, one can show that $f_{\alpha\beta\gamma}=0$  which in turn implies
that
$\cL(K)^{(1,0)}$ and  $\cL(K)^{(0,1)}$ are subalgebras of $\cL(K)\otimes \CN$.
The equation \senine\ then implies that $\omega_{\alpha}^s$ is a Lie algebra
homomorphism from
$\cL(K)^{(1,0)}$ into $\cL(H)\otimes \CN$.
Finally to find  (2,0) sigma models with off-shell supersymmetry and target
space group manifolds,
we assume that the connection $A$ satisfies (in addition to the requirements
necessary for the existence of (2,0) models with on-shell supersymmetry) the
condition $\nabla{\hat
I}=0$ which implies all the remainning conditions for off-shell closure of the
algebra of (2,0)
supersymmetry and Nijenhuis  transformations.  The condition
$\nabla{\hat I}=0$ has solutions provided the representation of $\cL(H)$ on the
fields $\psi$ has
an invariant (constant) complex structure and the fibre metric $h$ is hermitian
with respect to
${\hat I}$.


\chapter{The Poisson bracket algebra of charges}

The conserved currents of the (2,0) supersymmetric sigma model are the energy
momentum tensor, the (1,0) and (2,0) supersymmetry currents and a current
corresponding
to a $U(1)$ charge that rotates the two supersymmetry charges.   These currents
expressed in terms of (1,0) superfields are the following: The current
$$
{\cal G}^{{}^0}_{\pp +}=g_{ij} D_+\phi^i \partial_\pp\phi^j-{i\over 3}
H_{ijk}D_+\phi^i
D_+\phi^j D_+\phi^k \,
\eqn\done
$$
that has components the (1,0) supersymmetry current ${\cal S}^{{}^0}_{\pp
+}={\cal
G}^{{}^0}_{\pp +}|$ and the ${\cal T}_{\pp\pp}$ component of the energy
momentum
tensor, ${\cal T}_{\pp\pp}=-i{1\over2}\big(D_+{\cal G}^{{}^0}_{\pp +}\big)|$,
the
current
$$
{\cal G}^1_{+ +}= -I_{ij} D_+\phi^i  D_+\phi^j\ ,
\eqn\dtwo
$$
that has components the U(1) current ${\cal J}_{\pp}={\cal G}^1_{+ +}|$ and the
(2,0) supersymmetry current  ${\cal S}^{1}_{\pp +}=i{1\over2} D_+{\cal
G}^{1}_{\pp +}|$
and the current
$$
{\cal P}_{+\pp}={2\over 3}N_{ijk} D_+\phi^i D_+\phi^j D_+\phi^k
\eqn\dthree
$$
that has components the Nijenhuis currents ${\cal N}_{+\pp}=({\cal P}_{+\pp})|$
and
${\cal N}_{\pp\pp}=-{1\over4}(D_+{\cal P}_{+\pp})|$, where the
vertical line denotes the evaluation of the corresponding expression at
$\theta^+=0$.
All the above currents are chiral, i.e. $\partial_={\cal G}^{{}^0}_{+\pp}=0$,
$\partial_={\cal G}^{1}_{\pp}=0$ and $\partial_={\cal P}_{+\pp}=0$.

The most convenient way to present the Poisson bracket algebra of the charges
of the above currents
is to `smear' them with the parameters of the associated transformations.  The
charges are then the following:
$$
S^{{}^0}(\epsilon_=)=\int dx\, d\theta^+\, \epsilon_=\, {\cal G}^{{}^0}_{\pp
+}\ ,
\eqn\dfour
$$
$$
S^{1}(a_-)=\int dx\, d\theta^+\, a_-\, {\cal G}^{1}_{\pp}\ ,
\eqn\dfive
$$
and
$$
N(a_=)= \int dx d\theta^+\, a_=\, {\cal P}_{\pp +}\ .
\eqn\dsix
$$
It is worth pointing out that only certain linear combinations of all the
possible charges of
the model enter in the above expressions.  In particular,
$S^{{}^0}(\epsilon_=)$ is a linear
combination of the charges $T_\pp\equiv E+P$ and (1,0) supersymmetry charge  as
one can easily
verify by observing that
$$
S^{{}^0}(\epsilon_=)=(D_+\epsilon_=)|\int\  dx\  {\cal S}^{{}^0}_{\pp +}+ 2
i\epsilon_=| \int\  dx\  {\cal T}_{\pp \pp}\ ,
\eqn\dseven
$$
and $\epsilon_==\epsilon_=(\theta^+)$, where ${\cal T}_{\pp \pp}(\equiv
-i{1\over2} D_+{\cal
S}{{}^0}_{+\pp}|)$ is the indicated component of the energy momentum of the
theory.
Similarly,
$S^{1}(a_-)$ is proposional to (2,0) supersymmetry charge and
$N(a_=)$ is proposional to the charge of the ${\cal N}_{\pp\pp}$ Nijenhuis
current since
both parameters $a_-$ and $a_=$ are constant, {\sl i.e.} independent of the
co-ordinates of
the (1,0) superspace.  These charges can be easily expressed in terms of the
component fields
$\phi=\phi|$ and
$\lambda_+=(D_+\phi)|$ of the theory by performing the integration over the odd
variable $\theta$ and then substitute in the resulting expression the component
fields.

The non-vanishing Poisson brackets  of the charges \dfour-\dsix\ of the (2,0)
supersymmetric
sigma model with Nijenhuis symmetries are the following:
$$
\{S^{{}^0}(\epsilon_=),S^{{}^0}(\epsilon'_=)\}=
S^{{}^0}(iD_+\epsilon_=D_+\epsilon'_=) , \quad
\{S^1(a_-),S^1(a'_-)\}= -4 i S^{{}^0}(a_- a'_-)-2 N(a_- a'_-)\ .
 \eqn\deight
$$
This Poisson bracket algebra of the charges of
the (2,0)-supersymmetric sigma model with Nijenhuis symmetries is not
isomorphic to the
standard (2,0) supersymmetry algebra.  Indeed to compare the (2,0)
supersymmetry algebra
\deight\ and the corresponding (2,0) supersymmetry algebra  \inone , let us set
$$
S^{{}^0}(\epsilon_=)\equiv (D_+\epsilon_=)| S^{{}^0}_++2i\epsilon_=| T_\pp,
\qquad
 S^1(a_-)\equiv 2 ia_- S^1_+,
\qquad N(a_=)\equiv -4 a_= N_\pp
\eqn\deighta
$$
The Poisson bracket algebra \deight\ then becomes
$$
\{S^{{}^0}_+, S^{{}^0}_+\}=2T_\pp\ , \qquad \{S^{1}_+,
S^{1}_+\}=2(T_\pp+N_\pp)\ ;
\eqn\deightb
$$
the remaining Poisson brackets vanish.
Next we define new generators as follows:
$$
{\tilde S}^{{}^0}_+=S^{{}^0}_++ S^{1}_+\ , \qquad {\tilde
S}^{1}_+=-S^{{}^0}_++S^{1}_+\ ,
\qquad {\tilde T}_\pp=2T_\pp+N_\pp\ .
\eqn\deightc
$$
In terms of these new charges, the algebra \deightb\ can be rewritten as
$$
 \{{\tilde S}^{{}^0}_+ , {\tilde S}^{{}^0}_+\}=2 {\tilde T}_\pp\ , \qquad
\{{\tilde S}^{1}_+,
{\tilde S}^{1}_+\}=2 {\tilde T}_\pp\ ,  \qquad \{{\tilde S}^{{}^0}_+, {\tilde
S}^1_+\}=2 N_\pp\ .
\eqn\deightd
$$
The algebra \deight\ rewritten as \deightd\ is not isomorphic to the
corresponding (2,0)
supersymmetry algebra \inone\ because the Poisson bracket of first
supersymmetry charge ${\tilde
S}^{{}^0}_+$ with the second ${\tilde S}^{1}_+$ does not vanish  as in \inone\
but
rather it gives a new (central) charge
$N_\pp$ of the algebra which has Lorentz weight one.  Another feature of the
algebra
\deightd\ is that the $SO(2)$ rotation that rotates the supersymmetry charges
${\tilde
S}^{{}^0}_+$ and ${\tilde S}^{1}_+$ to each other and leaves the rest of the
charges
invariant is {\sl not} an automorphism of the algebra.  However the algebra
\deightd\ has an
$SO(1,1)$ (non-compact) automorphism $R$ that acts on its generators as
follows:
$$
\eqalign{
\{R,{\tilde S}^{{}^0}_+ \}&={\tilde S}^{1}_+\ ,\qquad \{R,{\tilde S}^1_+
\}={\tilde S}^{{}^0}_+\ ,
\cr
\{R,T_\pp \}&=2 N_\pp\ , \qquad \{R,N_\pp \}=2 T_\pp\ .}
\eqn\deighte
$$
This automorphism of the algebra is not realised by a transformation on the
fields of the
(2,0)-supersymmetric sigma model studied in section three.

To compute the Poisson bracket algebra of charges of (2,2)-supersymmetric sigma
models with Nijenhuis symmetries, section 2, it is convenient to express all
the charges in
terms of (1,1) superfields.  The currents of the theory are
$$\eqalign{
{\cal G}^{{}^0}_{+\pp}&= g_{ij} D_+\phi^i  \partial_\pp\phi^j-{i\over 3}
H_{ijk}
D_+\phi^iD_+\phi^jD_+\phi^k\ ,
\cr
{\cal G}^{{}^0}_{-=}&=   g_{ij} D_-\phi^i  \partial_=\phi^j-{i\over 3} H_{ijk}
D_-\phi^iD_-\phi^jD_-\phi^k\ ,
\cr
{\cal G}^{1}_{++}&= -I_{ij} D_+\phi^i D_+\phi^j\ ,
\cr
{\cal G}^{1}_{--}&= -J_{ij} D_-\phi^i D_-\phi^j\ ,
\cr
{\cal P}_{+\pp}&= {2\over3} N_{ijk} D_+\phi^i  D_+\phi^j D_+\phi^k\ ,
\cr
{\cal P}_{-=}&={2\over3} N_{ijk} D_-\phi^i D_-\phi^j D_-\phi^k\ ,}
\eqn\dnine
$$
where the components of the ${\cal G}^{{}^0}_{+\pp}$, ${\cal G}^{{}^0}_{-=}$
are the energy
momentum tensor and the (1,0) and (0,1) supersymmetry currents, the components
of ${\cal
G}^{1}_{++}$ and ${\cal G}^{1}_{--}$ are the (2,0) and
(0,2) supersymmetry currents and two $U(1)$ currents, and the components of
${\cal
P}_{+\pp}$ and ${\cal P}_{-=}$ are the currents corresponding to Nijenhuis
symmetries.  All the
above currents are conserved, {\sl i.e.}  $D_-{\cal G}^{{}^0}_{++}=0$,
$D_+{\cal
G}^{{}^0}_{-=}=0$, $D_-{\cal G}^{1}_{++}=0$, $D_+{\cal G}^{1}_{--}=0$,
$D_-{\cal P}_{+\pp}=0$
and $D_+{\cal P}_{-=}=0$. The associated `smeared' charges are as follows:
$$\eqalign{
S^{{}^0}(\epsilon_=)&=\int dx\, d\theta^+\, \epsilon_=\,  {\cal
G}^{{}^0}_{+\pp}\ ,
\qquad
S^{{}^0}(\epsilon_\pp)=\int dx\, d\theta^-\, \epsilon_\pp\,  {\cal
G}^{{}^0}_{-=}\ ,
\cr
S^{1}(a_-)&=\int dx\, d\theta^+\,  a_-\, {\cal G}^{1}_{++}\ ,
\qquad
S^{1}(a_+)=\int dx\, d\theta^-\,  a_+\, {\cal G}^{1}_{--}\ ,
\cr
N(a_=)&= \int dx\, d\theta^+\, a_=\, {\cal P}_{+\pp}\ ,
\qquad
N(a_\pp)= \int dx\, d\theta^-\, a_\pp\, {\cal P}_{-=}\ .}
\eqn\dten
$$
The non-vanishing Poisson brackets of the charges \dten\ of the
(2,2)-supersymmetric sigma
model are as follows:
$$ \eqalign{
\{S^{{}^0}(\epsilon_=),S^{{}^0}(\epsilon'_=)\}&=
S^{{}^0}(iD_+\epsilon_=D_+\epsilon'_=) , \quad
\{S^1(a_-),S^1(a'_-)\}= -4 i S^{{}^0}(a_- a'_-)-2 N(a_- a'_-) ,
\cr
\{S^{{}^0}(\epsilon_\pp),S^{{}^0}(\epsilon'_\pp)\}&= S^{{}^0}(iD_-\epsilon_\pp
D_-\epsilon'_\pp) ,
\quad
\{S^1(a_+),S^1(a'_+)\}= -4 i S^{{}^0}(a_+ a'_+)-2 N(a_+ a'_+)\ .}
\eqn\seeighteen
$$
The algebra of charges \seeighteen\ of the
(2,2)-supersymmetric sigma model is two commuting copies of the algebra
\deight\ of the (2,0)
model.  Using arguments similar to those for the (2,0) case, we can show that
\seeighteen\ is
not isomorphic to the corresponding (2,2) supersymmetry algebra \inone.


\chapter {Anomalies}

Generic sigma models with (2,0) supersymmetry have  a different number of left-
from right-
chiral fermions and therefore some of their symmetries are quantum mechanically
anomalous
[\sen, \hpc]. To examine the anomalies for (2,0)-supersymmetric sigma models
with
Nijenhuis symmetries, we quantise the theory in the background field method.
As in the
case of (2,0)-supersymmetric sigma models with target spaces complex manifolds,
the model can be
quantised in such a way that the  background/quantum field split symmetry
[\hps], (1,0)
supersymmetry and sigma model manifold reparameterisations are manifestly
preserved quantum
mechanically.  The arguments for this are similar to those of ref. [\hpc] and
they will not be
repeated here.  The transformations that may be anomalous quantum mechanically
are the following:
The frame rotations of the tangent bundle of the sigma model manifold
$$
\delta_U\omega^{(-)}_i{}^A{}_B=-\partial_iU^A{}_B+ U^A{}_C{}
\omega^{(-)}_i{}^C{}_B-\omega^{(-)}_i{}^A{}_C {} U^C{}_B \ ,
\eqn\aanone
$$
where $\omega^{(-)}$ is a spin connection of $\Gamma^{(-)}$ and $U$ is the
infinitesimal gauge
parameter,  the gauge transformations of the connection $A$
$$
\delta_LA_i{}^a{}_b=-\partial_iL^a{}_b+ L^a{}_c A_i{}^c{}_b -A_i{}^a{}_c
L^c{}_b\ ,
\eqn\aantwo
$$
where $L$ is the infinitesimal parameter,  the (2,0) supersymmetry
transformations
$$
\delta_I\phi^i=a_- I^i{}_jD_+\phi^j\ ,
\eqn\aanthree
$$
and the Nijenhuis symmetries
$$
\delta_N\phi^i=a_= N^i{}_{jk}D_+\phi^j D_+\phi^k\ ,
\eqn\aanfour
$$
where $N$ is the Nijenhuis tensor of the almost complex structure $I$.  The
transformations \aanone\ and \aantwo\ are invariances of the sigma model
action, {\sl i.e.}
transformations of the fields that leave invariant the classical action
provided that they are
compensated by appropriate transformations of the couplings\foot{Note that the
equations
\aanone\ and \aantwo\ denote only the transformations induced on the couplings
by these
invariances.}, and no Noether currents are associated with them.
 Because the (1,0) supersymmetry is manifestly preserved in the quantum theory
one can use (1,0)
superfields to study the anomalies of the classical symmetries
\aanone-\aanfour.

The anomalies in the frame rotations of the sigma model tangent bundle can be
derived from the
familiar descent equations [\bz]
$$
P_4=dQ_3^0\ , \qquad  \delta_U Q_3^0+dQ_2^1=0\ , \qquad \delta_U
Q_2^1+dQ_1^2=0\ ,\qquad \delta_U
Q_1^2+ dQ^3_0=0\ ,
\eqn\antwoa
$$
where $P_4$ is the first Pontrjagin class of the tangent bundle of the sigma
model target
space, $P_4={\rm tr} R^2$, and $U$ is the infinitesimal parameter of the
transformations.
Similar descent equations can be used to compute the anomalies of the gauge
transformations of
the connection $A$. The anomalies in the frame rotations of the sigma model
tangent bundle [\hw]
are
$$
\Delta(U)= i y \int d^2x d\theta^+\, Q_2^1\big(U, \omega^{(-)}\big)_{ij}
D_+\phi^i \partial_=\phi^j\ ,
\eqn\anone
$$
where
$$
Q_2^1(U, \omega^{(-)})= U^A{}_B(\phi) d\omega^{(-)}{}^B{}_A \ ,
\eqn\anonea
$$
and $y={{\hbar}\over 2\pi}$  is a numerical coefficient computed in
perturbation theory [\ht], and
similarly the anomalies in the gauge transformations of the connection
$A$ are
$$
\Delta(L)=-i y\int d^2x d\theta^+\, Q_2^1(L, A)_{ij} D_+\phi^i
\partial_=\phi^j\ ,
\eqn\antwo
$$
where $L$ is the infinitesimal parameter of the gauge transformations. The
connections
$\omega^{(-)}$ and $A$ that enter in the expressions for the anomalies are not
uniquely
specified by the descent equations.  In fact the anomalies \anone\ and \antwo\
can be
expressed in terms of any connection of the tangent bundle of $\cM$ and $\cE$,
respectively, by adding appropriate finite local counterterms in the effective
action.
However as we will see below in the supersymmetric case it is convenient to
express the
anomalies \anone\ and \antwo\ as above.  The rest of the anomalies are
specified by consistency
conditions. One can derive these consistency conditions by applying the
commutator of two
symmetries on the effective action of the quantum theory of a model and then
define the
non-vanishing variations of the effective action as the corresponding
anomalies.  The consistency
conditions for the (2,0)-supersymmetric sigma model with Nijenhuis symmetries
are the following:
$$
\delta_I\Delta(U)-\delta_U\Delta_I(a_-)=0\ , \qquad
\delta_I\Delta(L)-\delta_L\Delta_I(a_-)=0\ ,
\eqn\anthree
$$
$$\eqalign{
\delta_I\Delta_N(a_=)-\delta_N\Delta_I(a_-)&=0\ ,
\cr
\delta_N\Delta(U)-\delta_U\Delta_N(a_=)&=0\ ,
\cr
\delta_N\Delta(L)-\delta_L\Delta_N(a_=)&=0\ ,
\cr
\delta_N\Delta_N(a'_=)-\delta'_N\Delta_N(a_=)&=0\ ,}
\eqn\anfour
$$
and
$$
\delta_I\Delta_I(a'_-)-\delta'_I\Delta_I(a_-)=\Delta_N(a_-a'_-)\ ,
\eqn\anfive
$$
where $\Delta_I$ is the (2,0) supersymmetry anomaly and $\Delta_N$ is the
anomaly of the
Nijenhuis symmetry. Solving \anthree\ for the (2,0)
supersymmetry anomaly $\Delta_I$, we get
$$
\Delta_I(a_-)=3 y \int d^2x d\theta^+\, \big(
Q^0_3(\omega^{(-)},A)\big)_{ijk}\,
\delta_I\phi^i D_+\phi^j \partial_=\phi^k\ ,
\eqn\ansix
$$
where
$$
Q^0_3(\omega^{(-)})= {\rm tr}[\omega^{(-)}d\omega^{(-)}+{2\over 3}
\big(\omega^{(-)}\big)^3]
\eqn\ansixa
$$
is the Chern-Simons form of the connection $\omega^{(-)}$ and similarly  for
the
Chern-Simons form $Q^0_3(A)$ for the connection $A$, and
$$
Q^0_3(\omega^{(-)},A)=Q^0_3(\omega^{(-)})-Q^0_3(A)\ .
\eqn\ansixaa
$$
In fact \anthree\ specifies the (2,0)
supersymmetry anomaly up to a term invariant under both frame rotations and
gauge
transformations.  A direct computation of the one-loop effective action reveals
that such a term
does not appear. From the consistency condition \anfive, we get that the
anomaly of the Nijenhuis
symmetry is the following:
$$
\Delta_N(a_=)=3 y \int d^2x d\theta^+\, \big(
Q^0_3(\omega^{(-)},A)\big)_{ijk}\,
\delta_N\phi^i D_+\phi^j \partial_=\phi^k\ .
\eqn\anseven
$$
To prove this we have used that the curvature of the connections $\omega^{(-)}$
and $A$ are
(1,1) forms with respect to the almost complex structure $I$.  The rest of the
consistency
conditions \anthree\ and \anfour\ are also satisfied due to eqns.  \beight\ and
\bnine\ of
section three and similar equations satisfied by the curvature $R^{(-)}$ of the
$\omega^{(-)}$ connection.  The latter follows from the equations
\aeleven-\athirteen\ of section
two that involve the Nijenhuis tensor of the almost complex structure $I$ and
$R^{(-)}_{ijkl}=R^{(+)}_{klij}$.

To discuss a possible cancellation of the anomalies \anone, \antwo, \ansix\ and
\anseven, we
will briefly review the case that  $\cM$ is a complex manifold and $\cE$ is a
holomorphic vector bundle over $\cM$ [\hpc], {\sl i.e.}
there are not Nijenhuis symmetries in the theory.  In this case, one can
introduce complex
co-ordinates on  $\cM$ and an exterior derivation
$$
d_I=i (\partial-{\bar {\partial}})
\eqn\aneight
$$
where $\partial$ and ${\bar {\partial}}$ are the exterior derivatives with
respect the
holomorphic and anti-holomorphic co-ordinates of $\cM$, respectively. The
exterior
derivation $d_I$ is associated with the complex structure $I$ and satisfies the
following
conditions:
$$
d^2_I=0\ , \qquad dd_I+d_Id=0\ .
\eqn\annine
$$
Using \annine, the Poincar{\'e} lemma, the Dolbeault-Grothendieck lemma and the
fact that
$$
P_4(\omega^{(-)},A)\equiv P_4(\omega^{(-)})- P_4(A)
\eqn\anninea
$$
is a (2,2)-form on $\cM$ with respect to the complex structure $I$, one can
write the three-form
$Q_3^0(\omega^{(-)},A)$ as follows:
$$
Q_3^0(\omega^{(-)},A)=dX+d_IY\ ,
\eqn\anten
$$
where $X$ is a two-form and $Y$ is a (1,1)-form with respect to $I$ on $\cM$.
Next
the finite local counterterm
$$
\Gamma_{fl}=-i y \int \, dx\, d\theta^+\, (X_{ij}-Y_{ik} I^k{}_j) D_+\phi^i
\partial_=\phi^j
\eqn\aneleven
$$
 cancels the anomalies \anone, \antwo\ and
\ansix\ [\hpc].  However the finite local counterterm \aneleven\ depends
explicitly on the
holomorphic structure of $\cM$ and $\cE$ and therefore induces new anomalies in
the holomorphic
gauge transformations of the tangent bundle of $\cM$ and $\cE$. The holomorphic
anomalies are
cancelled by anomalous variations of the metric and Wess-Zumino term of the
theory.  Finally,
there is global anomaly in the theory which cancels provided that the four-form
\anninea\ is
exact [\nel].

Now we will turn our attention to the case that the almost complex structure
$I$ is not
integrable. It is well known that for every vector valued form on
$\cM$ one can introduce an exterior derivation.  In particular there are
exterior
derivatives $d_I$ and $d_N$ associated with the almost complex structure $I$
and the Nijenhuis
tensor $N$ of $I$, respectively (see for example ref.[\hpb]). One can show that
the derivations $d,
d_I$ and
$d_N$  obey the following  algebra:
$$\eqalign{
d^2=&0\ , \qquad 2d_I^2=d_N\ ,\qquad  dd_I+d_Id=0\ ,
\cr
dd_N&-d_Nd=0\ , \qquad d_Id_N-d_Nd_I=0\ .}
\eqn\antwelve
$$
Therefore in this case the derivation $d_I$ is not nilpotent and there is no
analogue of the
Dolbeault-Grothendieck lemma that it is necessary to write
$Q_3^0(\omega^{(-)},A)$ as in \anten.
So for generic sigma models with target spaces almost complex manifolds, the
(2,0) supersymmetry
and Nijenhuis transformations are anomalous quantum mechanically.  However the
anomalies due to
frame rotations of the tangent bundle of $\cM$ and the gauge transformations of
$A$ still cancel by
anomalous variation of the Wess-Zumino term of the theory as in the case of
(1,0)-supersymmetric
sigma models in refs. [\hw, \ht].  In some special cases of sigma models with
Nijenhuis symmetries
the (2,0) supersymmetry and Nijenhuis  anomalies cancel as well. Indeed if the
four-form
$P_4(\omega^{(-)},A)$ of eqn.
\anninea\ can be chosen to be zero, then $Q_3^0(\omega^{(-)},A)=dZ$ and the
finite local
counterterm
$$
\Gamma_{fl}=-i y \int \, dx\, d\theta^+\, Z_{ij} D_+\phi^i \partial_=\phi^j
\eqn\aneleven
$$
can be added in the effective action of the theory such that the anomalies in
the
(2,0) supersymmetry  and Nijenhuis transformations vanish. Examples of such
(2,0)-supersymmetric
sigma models are those studied in section four with target spaces  group
manifolds. For these
models
$\omega^{(-)}=0$ and one can choose $A=0$ so all the anomalies \anone, \antwo,
\ansix\ and
\anseven\ vanish identically.

Another example is the anomaly cancellation in the case of (2,2)-supersymmetric
sigma model.
This model is not expected to be anomalous because it has equal number of left
and right handed
fermions.  One of the conditions for a (2,0)-supersymmetric sigma model to be
(2,2)-supersymmetric is to set
$A=\Gamma^{(-)}$ in which case $P_4(\omega^{(-)},A)=0$ and this condition is
sufficient for the
cancellation of (2,0) supersymmetry \ansix\ and  Nijenhuis \anseven\ anomalies.
 A
similar argument can be used to prove that the (0,2) supersymmetry anomaly and
associated
Nijenhuis one cancel as well.



\chapter{Concluding Remarks}
\section {Massive (2,0) models}

The results of section three on massless (2,0)-supersymmetric sigma models with
target
spaces almost complex manifolds can be extended to massive ones.  For this, the
action of massive sigma models with (1,0) supersymmetry [\hpt,
\pt] is the following:
$$
I_m= -i \int\, dx\, d\theta^+ \{ (g+b)_{ij} D_+\phi^i \partial_=\phi^j+ i
h_{ab}
\phi^a_-\nabla_+\psi^b_-+i m s_a(\phi) \psi^a_-\}
\eqn\mone
$$
where the fields $\phi$, $\psi$ and the couplings $g,b,A$ are defined as in
section three.  The
only new coupling is $s_a$ which can be thought as a section of the bundle
$\cE$ over $\cM$ and
$m$ is a mass parameter.  The action \mone\ is manifestly (1,0) supersymmetric.
 The (2,0)
supersymmetry transformations written in (1,0) superfields are
$$\eqalign{
\delta_I\phi^i&=a_- I^i{}_j D_+\phi^j\ ,
\cr
\delta_I\psi^a&={-i a_-} \hat{I}^a{}_b {\cal{S}}^b +{1\over2} m a_- t^a(\phi)}
\eqn\mtwo
$$
where $I$ is a (1,1) tensor of $\cM$, $\hat{I}$ is a tensor on the fibre of
$\cE$, $t^a$ is a
section of $\cE$ and
$$
{\cal{S}}^a\equiv 2 i \nabla_+\psi^a_-+im s^a
\eqn\mthree
$$
is the field equation for $\psi$. The transformations \mtwo\ leave the action
\mone\ invariant
and close on-shell to translations and Nijenhuis transformations,
$$\eqalign{
\delta_{N}\phi^i&=a_= N^i_{jk} D_+\phi^j D_+\phi^k\ ,
\cr
\delta_{N}\psi_-^a&=- A_i{}^a{}_b \delta_{N}\phi^i \psi_-^b\ ,}
\eqn\mfour
$$
provided that, in addition to the conditions described in section three for the
invariance of the
action of the massless (2,0) model ($m=0$) and the on-shell closure of the
algebra of the
associated (2,0) supersymmetry transformations, the new condition
$$
\nabla_it^a-I^j{}_i \nabla_js^a=0
\eqn\mfive
$$
is satisfied.  The closure of the algebra of (2,0) supersymmetry
transformations \mthree\ has
been studied before for the special case where $I$ is a complex structure
[\hpt, \pt].
Finally, the Nijenhuis transformations \mfour\ leave the action invariant and
the algebra of
transformations \mtwo\ and \mfour\ closes provided that, in addition to the
conditions
\aeleven-\athirteen\ and \beight\ given in sections two and three, the
following conditions are
satisfied:
$$
N^k_{ij}\nabla_ks^a=0, \qquad N^k_{ij}\nabla_kt^a=0 \ .
\eqn\msix
$$
The conditions \msix\ are not independent but they are integrability
conditions for
\mfive\ and the condition that $F(A)$ is a (1,1)-form with respect to the
almost complex structure
$I$. The algebra of transformations \mtwo\ and \mfour\ of the massive
(2,0)-supersymmetric sigma
model is isomorphic to the algebra of transformations \bthree\ and \bfive\ of
the massless model.
Finally, it is straightforward to extend the above results to massive
(2,2)-supersymmetric sigma
models with Nijenhuis symmetries.


\section{Topological models}

One way to associate a topological sigma model to a supersymmetric one is by
twisting the
supersymmetry algebra of the latter [\witten].   The traditional way to twist
the supersymmetry
algebra of a sigma model is to define a new Lorentz generator which is the sum
of the original
Lorentz  generator of the theory with a linear combination of the generators of
the abelian
subgroup of the automorphism group of the its supersymmetry algebra that rotate
supersymmetry
charges but leave the rest of generators of the algebra invariant. Then new
Lorentz weights
for the generators of the supersymmetry algebra are assigned with respect to
this new Lorentz
generator.  Provided that the new Lorentz generator is chosen appropriately,
some of the
supersymmetry charges of the sigma model transform under the new Lorentz
transformations as
scalars and the Poisson brackets of certain linear combinations of them either
vanish or close to
a commuting  central charge of the algebra with new Lorentz weight zero. Such
charges that are
anticommuting, Lorentz scalars and their Poisson brackets either vanish or
close to a central
charge can be thought as BRST charges and these are the charges that generate
the symmetries of
the associated topological model.

The method described above  to associate a topological sigma model to a
supersymmetric one does not seem to be applicable in the case of
(2,0)-supersymmetric sigma models
with Nijenhuis symmetries because  rotations of the supersymmetry charges of
\intwo\ that leave
the rest of the charges invariant are not automorphisms of the algebra.
An alternative way to twist the algebra \intwo\ is to use the automorphism
\deighte\ and define a
new Lorentz generator $L'=L-{1\over2}R$. It is important though to note that
the automorphism $R$
is not realised by transformations on the fields of the (2,0)-supersymmetric
sigma model with
Nijenhuis symmetries and therefore the relation between the
(2,0)-supersymmetric sigma model and
its topological version is not as clear. Nevertheless, all the charges of the
(2,0) supersymmetry
algebra \deightb\ are Lorentz scalars with respect to $L'$ and the
non-vanishing Poisson brackets
of the twisted (2,0) supersymmetry algebra are
$$
\{S^{{}^0},S^{{}^0}\}=2 T, \qquad \{S^1,S^1\}=2 W, \qquad \{S^{{}^0},S^1\}=0\ ,
\eqn\tone
$$
where $S^{{}^0},S^1$ are anti-commuting charges and $T, W$, ($W=T+N$), are
commuting ones.   This
topological algebra is similar to the topological algebra that one gets in the
equivariant version
of topological sigma models, ref. [\witten], and similar methods can be applied
to construct
models. The above topological twisting of the (2,0) supersymmetry algebra can
be easily
generalised to twist (2,2) one.


\chapter{Summary}

Sigma models with $N=2$ supersymmetry have been studied extensively in the
literature  because of their applications to string compactifications,
integrable systems and the
novel renormalisation properties of some of their couplings. Most of the effort
so far, with some
exceptions, has been concentrated to investigate those $N=2$-supersymmetric
sigma models that
their  algebra of charges is isomorphic to the standard one \inone. These
models have target
spaces that are complex manifolds.  However  we have found a new class of
(2,0)-supersymmetric
two-dimensional sigma models with target spaces almost complex manifolds
extending similar
results for (2,2)-supersymmetric sigma models.  The supersymmetry algebras of
these (2,2)- and
(2,0)-supersymmetric sigma models close on finite number of generators and
apart from the
supersymmetry charges, the energy and momentum, they contain other charges of
Lorentz weight one
that generate new symmetries of the sigma model action associated to the
Nijenhuis tensor of the
almost complex structures.  We have shown that the supersymmetry algebras of
sigma models with
Nijenhuis symmetries are not isomorphic to standard one ( eqn. \inone).  The
supersymmetry algebras
of (2,2)- and (2,0)-supersymmetric sigma models with target space an almost
complex manifold are
{\sl not} a counter-example to the Haag-Lopuszanski-Sohnius theorem because
this theorem deals
with the structure of four-dimensional supersymmetry algebras.  Supersymmetry
algebras isomorphic
to the one of (2,2)- and (2,0)-supersymmetric sigma models with target spaces
almost complex
manifolds have been have been discussed before in the context of string theory.
In the string
case, however, the space-time is taken to be flat but topologically non-trivial
and the additional
Lorentz weight one central charges are associated to winding numbers [\pkt].
The massive (2,0)-
and (2,2)-supersymmetric sigma models with target spaces almost complex
manifolds have been also
considered.

The anomalies in the symmetries of the (2,0)-supersymmetric sigma models with
target space almost
complex manifolds have been investigated. It can be arranged for anomalies to
occur in
the frame rotations of the sigma model manifold, the gauge transformations of
connection coupling
of the fermionic sector, the (2,0)-supersymmetry and the Nijenhuis
transformations.  The cancellation of these anomalies has been studied, as
well, and it has been
found that some models are anomaly free like for example (2,0)-supersymmetric
sigma models on
group manifolds.

The automorphism group of the supersymmetry algebra of (2,0)- and
(2,2)-super\-sym\-metric  sigma
models with target space almost complex manifolds does not contain an element
that rotates the
supersymmetry charges but leaves the rest of the charges invariant in
contradistinction to the
automorphism group of the standard supersymmetry algebra \inone.  However it
has been found that
there is another automorphism that rotates the supersymmetry charges and the
two Lorentz weight
one charges.  This automorphism cannot be realised by field transformations of
the
associated sigma model but it can be used to twist the supersymmetry algebra.
The twisted
supersymmetry algebra is similar to the topological algebra that one gets in
the
context of equivariant topological sigma models.

All (2,0)- and (2,2)-supersymmetric massless sigma models with off-shell
supersymmetry
and algebra of charges isomorphic to \inone\  admit a manifest (2,0) and (2,2)
conventional
superfield formulation in terms of constrained superfields, respectively. A
superfield
formulation has also been constructed for some (2,2)-supersymmetric sigma
models  with on-shell
supersymmetry and target spaces complex manifolds [\mr].  We have shown that
there are off-shell
(2,0)-supersymmetric sigma models with target spaces almost complex manifolds.
However a
superfield formulation of the these models similar to that of ref. [\hpa] for
(2,0)-supersymmetric
sigma models with target spaces complex manifolds does not seem to be
applicable.  We leave to
the future the study of this problem.

\vskip 1.0cm

\vfill\eject

\noindent{\bf Acknowledgments:} I would like to thank Dieter L{\"u}st for an
invitation to
visit Humboldt University in Berlin where part of this paper was completed.
Thanks are also
due to Hermann Nicolai for an invitation to visit the University of Hamburg, at
DESY.  I am
grateful to Paul Howe, Christian Preitschopf, Paul Townsend and Anton van de
Ven for discussions.
I am supported by a University Research Fellowship from the Royal Society.

\refout

\bye